\documentclass[referee,traditabstract]{aa}
\usepackage{graphicx,amsmath,relsize,epsfig,epstopdf}
\usepackage[varg]{txfonts}

\begin{document}

\title{Coronal magnetic field strength from Type II radio emission: complementarity with Faraday rotation measurements}
\author{S. Mancuso \inst{1} \and M. V. Garzelli \inst{2}}
\institute{INAF - Astrophysical Observatory of Torino, Strada Osservatorio 20, Pino Torinese 10025, Italy \inst{1} \\
Laboratory for Astroparticle Physics, University of Nova Gorica, SI-5000 Nova Gorica, Slovenia \inst{2}}

\date{Received / Accepted}
\mail{$\,\,$ S. Mancuso: mancuso@oato.inaf.it, $\,\,$ M.V. Garzelli: garzelli@mi.infn.it .}

\abstract{We analyzed the band splitting of a Type~II radio burst observed on 1997 May 12 by ground- and space-based radio spectrometers. Type II radio emission is the most evident signature of coronal shock waves and the observed band splitting is generally interpreted as due to plasma emission from both upstream and downstream shock regions. From the inferred compression ratio  we estimated, using the magnetohydrodynamic (MHD) Rankine-Hugoniot relations, the ambient Alfv\'en Mach number. By means of the  electron density obtained by inverting white-light polarized brightness  ($pB$)  coronagraph data and the shock speed inferred from the Type~II frequency drift,  we finally derived a radial profile for the magnetic field strength in the middle corona. The result was compared with the field profile obtained in May 1997 (but above $\sim 5$ $R_{\sun}$) with Faraday rotation measurements of extragalactic radio sources occulted by the corona. The power law of the form $B(r) ~ = ~ 3.76 ~ r^{-2.29} { ~ ~ \rm G}$ inferred in that work nicely describes the combined set of data  in a wide range of heliocentric distances ($ r \approx 1.8 - 14$ $R_{\sun}$).
}
\titlerunning{Coronal magnetic field strength from Type II radio emission}
\authorrunning{Mancuso \& Garzelli}
\maketitle

\section{Introduction}

In a recent paper (Mancuso \& Garzelli 2013), we analyzed Faraday rotation measurements of extragalactic radio sources occulted by the corona in order to assess the inner heliospheric magnetic field in May 1997, shortly after the previous solar minimum in 1996. By inverting polarized brightness ($pB$) data taken from the Large Angle and Spectrometric Coronagraph (LASCO)  aboard the Solar and Heliospheric Observatory (SOHO)  during the days of observation, we were able to estimate the strength of the heliospheric magnetic field in a range of heliocentric distances spanning from $\sim 5~ R_{\sun}$ to 14 $ R_{\sun}$. The Faraday rotation measurements used in that paper, made by Mancuso \& Spangler (2000) with the Very Large Array (VLA) radio telescope of the National Radio Astronomy Observatory (NRAO), have been limited to heliocentric distances greater than 5 $R_{\sun}$ because of a reduction in sensitivity due to solar interference in the beam side lobes. However, field estimates at shorter heliocentric distances are of pivotal importance since the strength and structure of the magnetic field in the region $\lesssim 5 $ $R_{\sun}$ crucially affects the acceleration of the fast solar wind component, that in turn has strong influence on the space weather at larger distances.
Among the known techniques for measuring the magnetic field in the middle corona, suitable coronal diagnostics are provided by the analysis of the radiation emitted by the local plasma during the passage of coronal shock waves (e.g., Mann et al. 1995: Mancuso et al. 2002, 2003; Bemporad \& Mancuso 2010; Gopalswamy \& Yashiro 2011).  In the corona, when the difference in speed between an outward propagating coronal mass ejection (CME) and the solar wind is larger than the local  fast-mode speed $v_{\rm F}$\footnote{ $v_{\rm F}^2 \equiv {{1} \over {2}} \left[v_{\rm A}^2 + c_{\rm S}^2 + \sqrt{(v_{\rm A}^2 + c_{\rm S}^2)^2 - 4v_{\rm A}^2 c_{\rm S}^2\cos^2\theta_\mathrm{Bn}}\right]$, where  $v_{\rm A}$ is the Alfv\'en speed, $c_{\rm S}$ the sound speed, and $\theta_\mathrm{Bn}$ the angle between the magnetic field and the direction of propagation.}, a forward magnetohydrodynamic (MHD) shock is produced ahead of the front. The collisionless shock leads to the generation of irreversible dissipation processes in the plasma, resulting in heating, acceleration of  particles, generation of entropy, and emission of radiation. The accelerated electrons form velocity space beam distributions in the foreshock region that are unstable to the generation of Langmuir waves via wave--particle interactions, which in turn produce electromagnetic emission, leading to the formation of Type~II radio bursts in dynamic spectra. Type~II radio bursts appear as bands of enhanced radio emission slowly drifting from high to low frequencies and often show a typical fundamental-harmonic structure, i.e, two drifting bands with a frequency ratio about 2:1.
The fundamental  band, emitted at the electron plasma frequency $f_\mathrm{pe} \approx 9 \sqrt{n_\mathrm{e}[\rm cm^{-3}]}$ kHz, relates directly to the local electron density $n_\mathrm{e}$ and thus to the burst driver's height. The measured frequency drift rate $D_f$ at a given time is directly related to the shock speed  $v_\mathrm{sh}$ and thus provides information on the shock dynamics through the corona. Both observations and theories suggest that Type~II radio emission is generated just upstream of the shock (Cairns 1986; Bale et al. 1999), so that it should refer to ambient plasma rather than to shocked  material. In a few cases, however, Type~II emission appears to be split into two parallel lanes separated by a few MHz and is interpreted as plasma emission occurring both upstream and downstream of the shock front (e.g, Vr{\v s}nak et al. 2002; Cho et al. 2007; Mancuso \& Avetta 2008; Magdaleni{\'c} et al. 2010). Under appropriate assumptions, the application of the Rankine-Hugoniot jump conditions, which relate the observed  band splitting to the shock compression ratio, allows the local Alfv\'en Mach number $M_\mathrm{A}= u_\mathrm{sh}/v_\mathrm{A}$ to be inferred, where $u_\mathrm{sh}$ is the shock speed in the solar wind frame ($u_\mathrm{sh} = v_\mathrm{sh} - v_\mathrm{sw}$). Finally, since by definition $v_\mathrm{A} \approx 2\cdot10^{11} B/\sqrt{n_\mathrm{e}}$, if information on $n_\mathrm{e}(r)$ and $u_\mathrm{sh}$ are available, $B(r)$ can be easily estimated. In this Letter, by  analyzing  the band splitting of a metric Type~II radio burst detected on 1997 May 12, we will show that this technique can be complementary to the coronal Faraday rotation method and is thus  able to extend the profile of the magnetic field strength obtained at distances greater than 5 $R_{\sun}$ down to  about 2 $R_{\sun}$ . In Sect. 2 we describe the observations; in Sect. 3 we show the results. A summary and discussion are presented in Sect. 4.

\begin{figure}[t]
\centering
\includegraphics[width=0.9\textwidth]{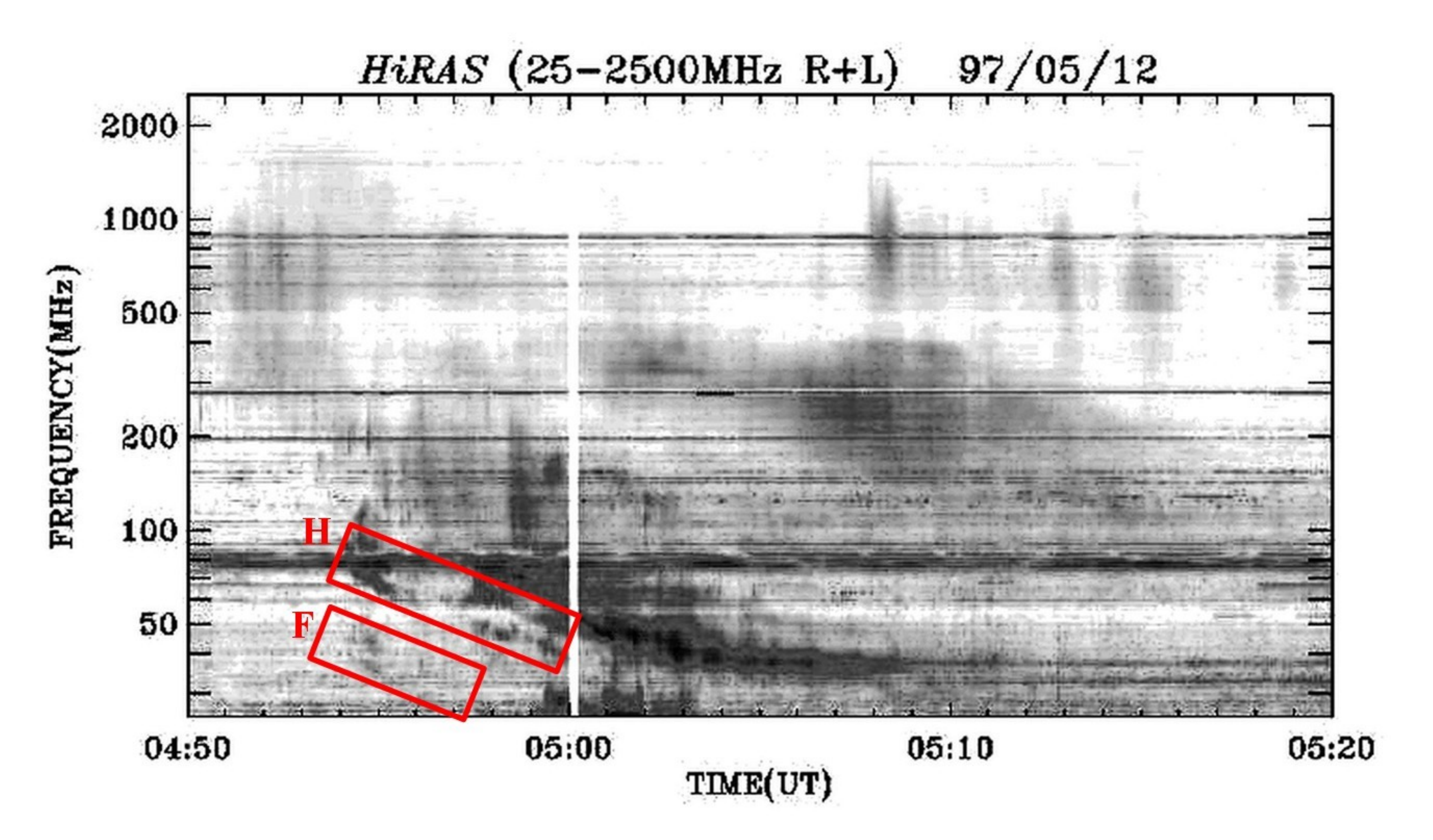}
\caption{Radio dynamic spectrum from the Hiraiso Radio Spectrograph (HiRAS;  inverse color scale: dark shading corresponds to bright emission) showing the Type~II burst observed on 1997 May 12.The superimposed red boxes highlight the fundamental (F) and harmonic (H) split bands under study.}
\end{figure}

\begin{figure}[h]
\centering
\includegraphics[width=0.9\textwidth]{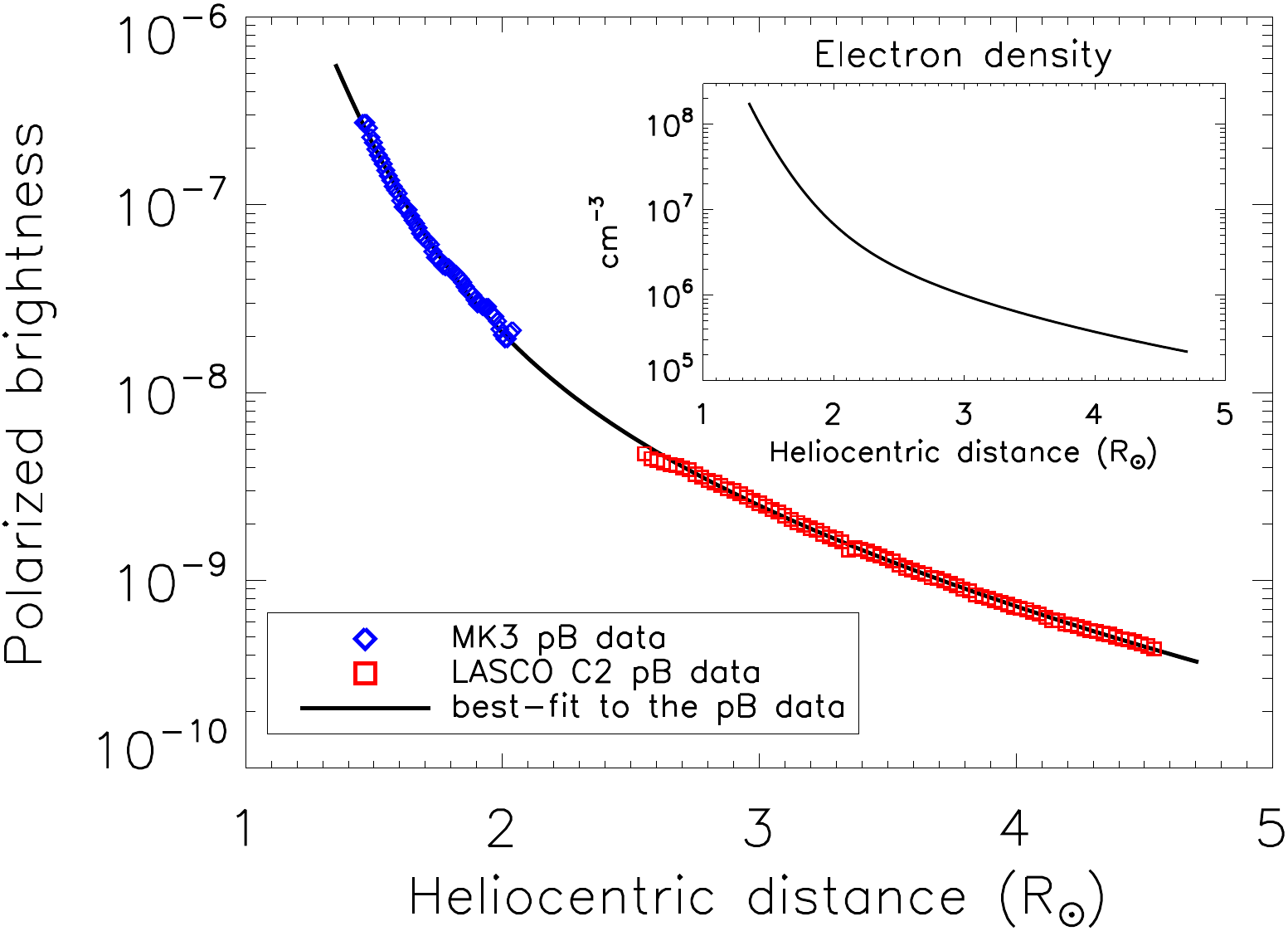}
\caption{$pB$ radial profile from Mk3 ({\it blue diamonds}) and LASCO C2 ({\it red squares}) observations obtained on 1997 May 5 above the solar equator at the east limb and radial power-law best-fit to the data ({\it solid black line}); the inset shows the electron density profile $n_\mathrm{e}(r)$ obtained from the $pB$ inversion procedure.}
\end{figure}

\section{Observations  and data reduction}

On 1997 May 12, a Type~II radio burst was observed by both ground- and space-based radio instruments; it was a consequence of a full-halo CME event detected by LASCO C2 on board SOHO.  The CME, expanding radially with an estimated speed of $\sim 600$ km s$^{-1}$  (Plunkett et al. 1998), originated near the center of the Sun's disk and was associated with an eruptive event observed by the Extreme Ultraviolet Imaging Telescope (EIT/SOHO) centered on active region NOAA 8038 at N20 W07. The metric Type~II burst observed by the Hiraiso Radio Spectrograph (HiRAS; Fig. 1) shows both fundamental and harmonic components starting at 4:54 UT, 12 min after the onset of the associated GOES C1.3 X-ray flare. The Type~II emission was also visible in the RAD2 dynamic spectrum of the Radio and Plasma Wave Experiment (WAVES) on board the Wind spacecraft  for a short while below 13.8 MHz, ending at $\sim 10$ MHz (Gopalswamy \& Kaiser 2002). Both Hiraiso and WAVES radio dynamic spectra show episodes of intermittent band splitting between about 4:54 - 5:00 UT in the metric band and around 5:12 UT in the decametric range. The two emission lanes show correlated frequency drifts, similar morphologies, and intensity variations, so that the emission is thought to originate from a common radio-source trajectory (Vr{\v s}nak et al. 2001) rather than from different portions of one global shock surface. We thus interpret the observed band-splitting episodes as a consequence of plasma radiation from the regions upstream and downstream of the shock. 

The estimate of the magnetic field strength from Type~II  band-splitting observations is particularly sensitive to the adopted density profile. Observations of the extended corona are primarily obtained with white-light coronagraphs that are able to detect coronal structures highlighted by photospheric radiation Thomson-scattered by free electrons in the ionized plasma.  In general, it is difficult to infer $n_\mathrm{e}(r)$ in the corona with remote-sensing techniques since the observed emission arises from the contribution of different structures integrated along the line of sight. However, during the previous solar minimum, the overall structure of the corona was highly axially symmetric so that $n_\mathrm{e}(r)$  can be confidently estimated from the $pB$ values extracted from the coronograph images by inverting a line-of-sight integral, according to the technique developed by van de Hulst (1950).
The strength of a MHD fast-mode shock in the corona can vary because of the inhomogeneous distribution of $v_\mathrm{A}$, but it is very much enhanced on those parts of the wave front that encounter low-$v_\mathrm{A}$ structures (Kahler \& Reames 2003;  Mancuso \& Raymond 2004). For this reason, we  applied the $pB$ inversion technique along the densest part of the streamer belt (the equatorial region). In this work, we use a combined set of $pB$ observations from both the NCAR/HAO Mauna Loa Solar Observatory's (MLSO) MK3 coronameter and LASCO C2 recorded on the east limb about a week before the CME onset (Fig. 2). We implicitly  assume that the corona has rotated almost rigidly during the interval of time from the reference $pB$ observations to the onset of the Type~II radio burst since no CME events have been recorded during this period above the east limb (according to the online SOHO/LASCO CME catalog available at the CDAW Data Center; Gopalswamy et al. 2009). From the combined $pB$ radial profile, a radial power-law fit of the form $pB(r) = Ar^{-\alpha} + Br^{-\beta}$ (with $r$ in units of $R_{\sun}$) was performed, after which a radial profile  $n_\mathrm{e}(r) = 4.27\cdot 10^9r^{-9.97} + 4.32\cdot 10^7r^{-3.25}$, roughly corresponding to three times the  Saito et al. (1977) density profile, was obtained\footnote{An enhancement factor of $C_n = 1.3$ was also applied as evinced from the analysis of Mancuso \& Garzelli (2013).  We note that the coronal density profile obtained in this work differs from that used in the previous analysis because the two profiles refer to different regions in the solar corona. The previous work concerned heliocentric distances up to $\sim 14$ $R_{\sun}$, where an $\sim r^{-2}$  dependence for $n_\mathrm{e}(r)$ is expected. For this work,  the radial dependence is certainly much steeper in the height range of interest ($< 3$ $R_{\sun}$). }.

\begin{figure}[t]
\centering
\includegraphics[width=0.9\textwidth]{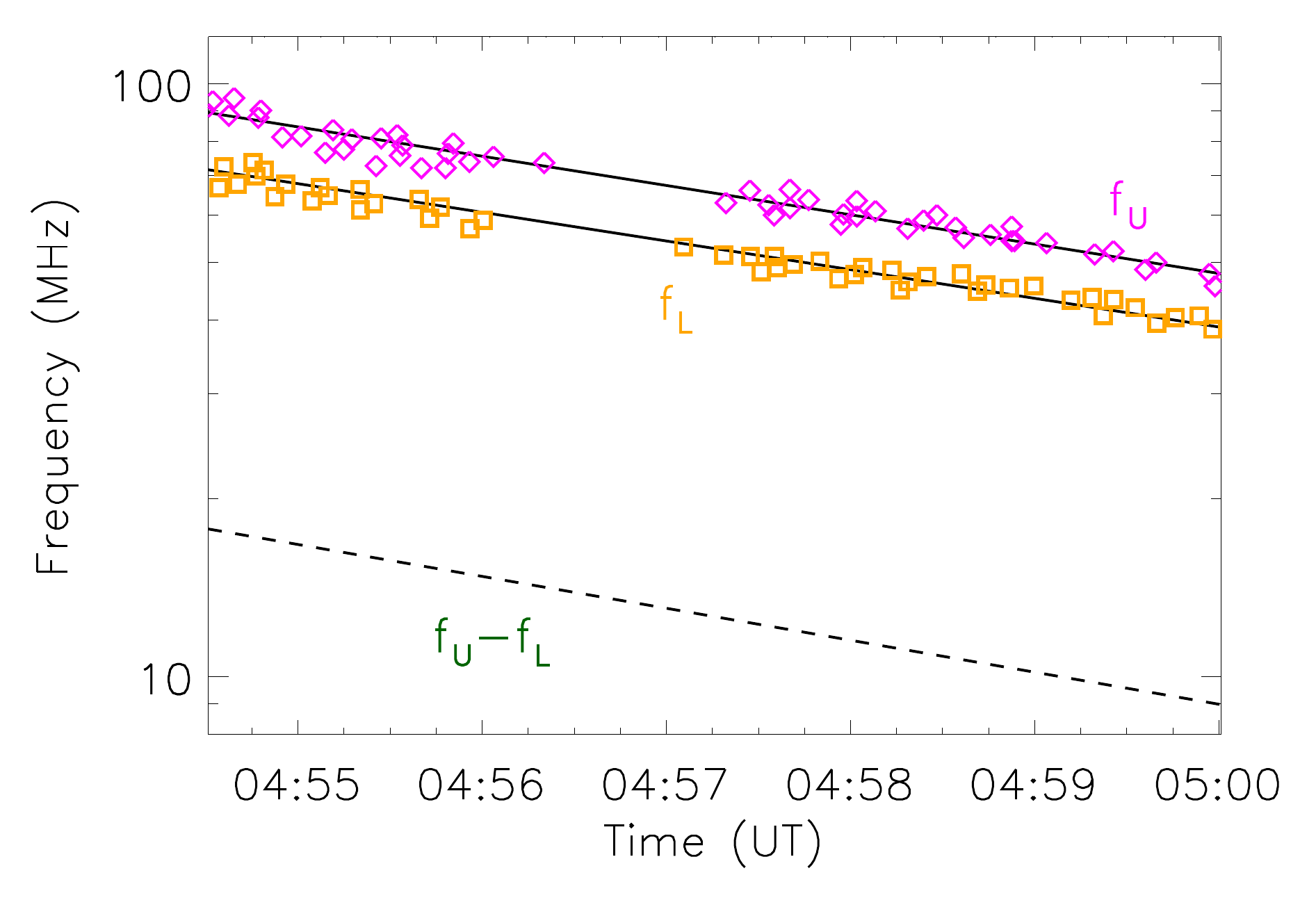}
\caption{Measured emission frequencies as a function of time as derived from the analysis of the upper ($f_\mathrm{U}$) and lower ($f_\mathrm{L}$) frequency branches of the harmonic band  of the 1997 May 12 Type~II burst. Solid black lines represent linear fits to the logarithmic data; the dashed line shows the difference ($f_\mathrm{U} - f_\mathrm{L}$) of the fitted lines.}
\end{figure}
\begin{figure}[b]
\centering
\includegraphics[width=0.9\textwidth]{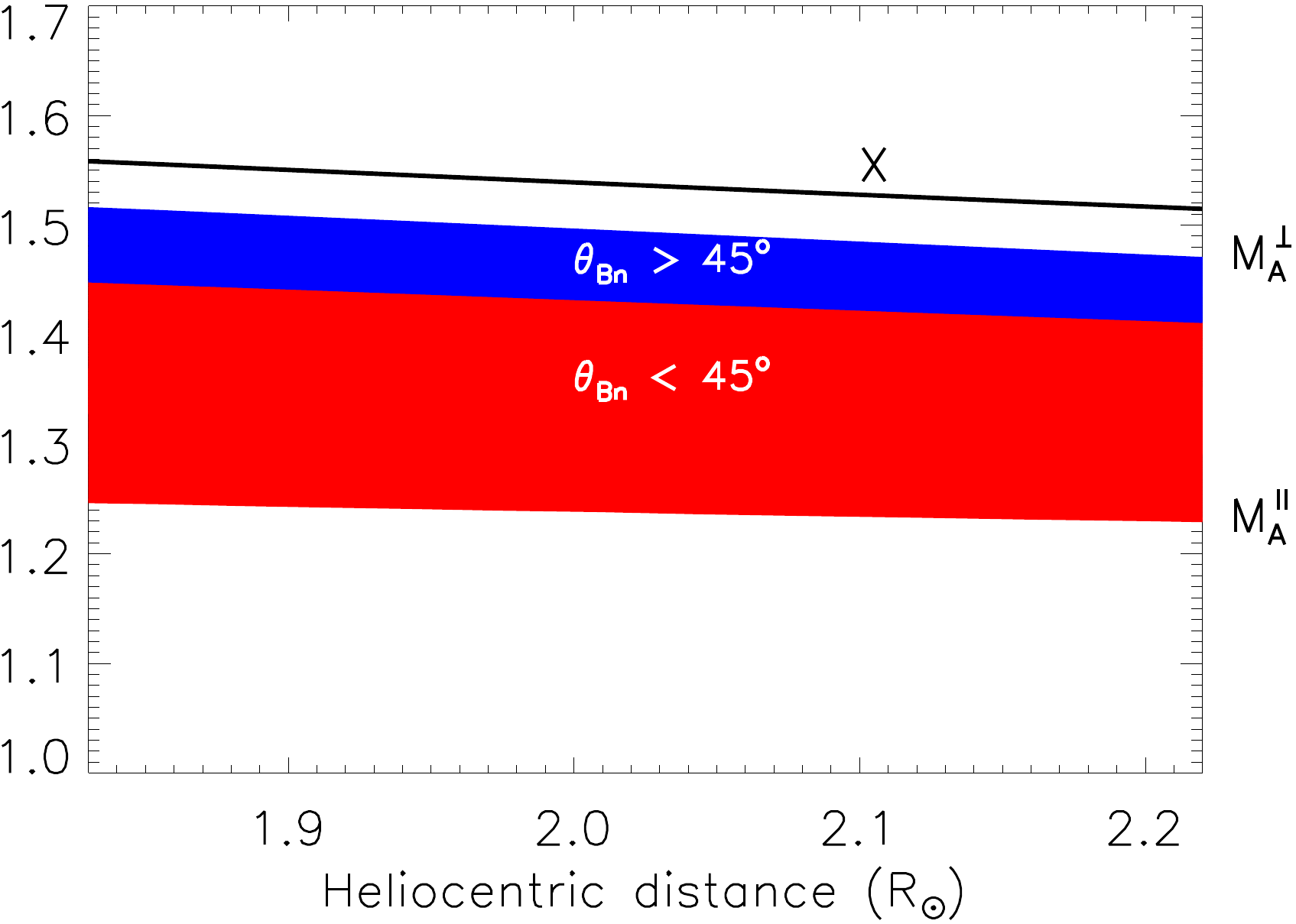}
\caption{Compression ratio $X$ ({\it black line}) and Alfv\'en Mach number $M_\mathrm{A}$ for quasi-perpendicular ({\it blue area}) and quasi-parallel ({\it red area}) cases as obtained  from the band-splitting structure observed in HiRAS.}
\end{figure}
\begin{figure}[t]
\centering
\includegraphics[width=0.9\textwidth]{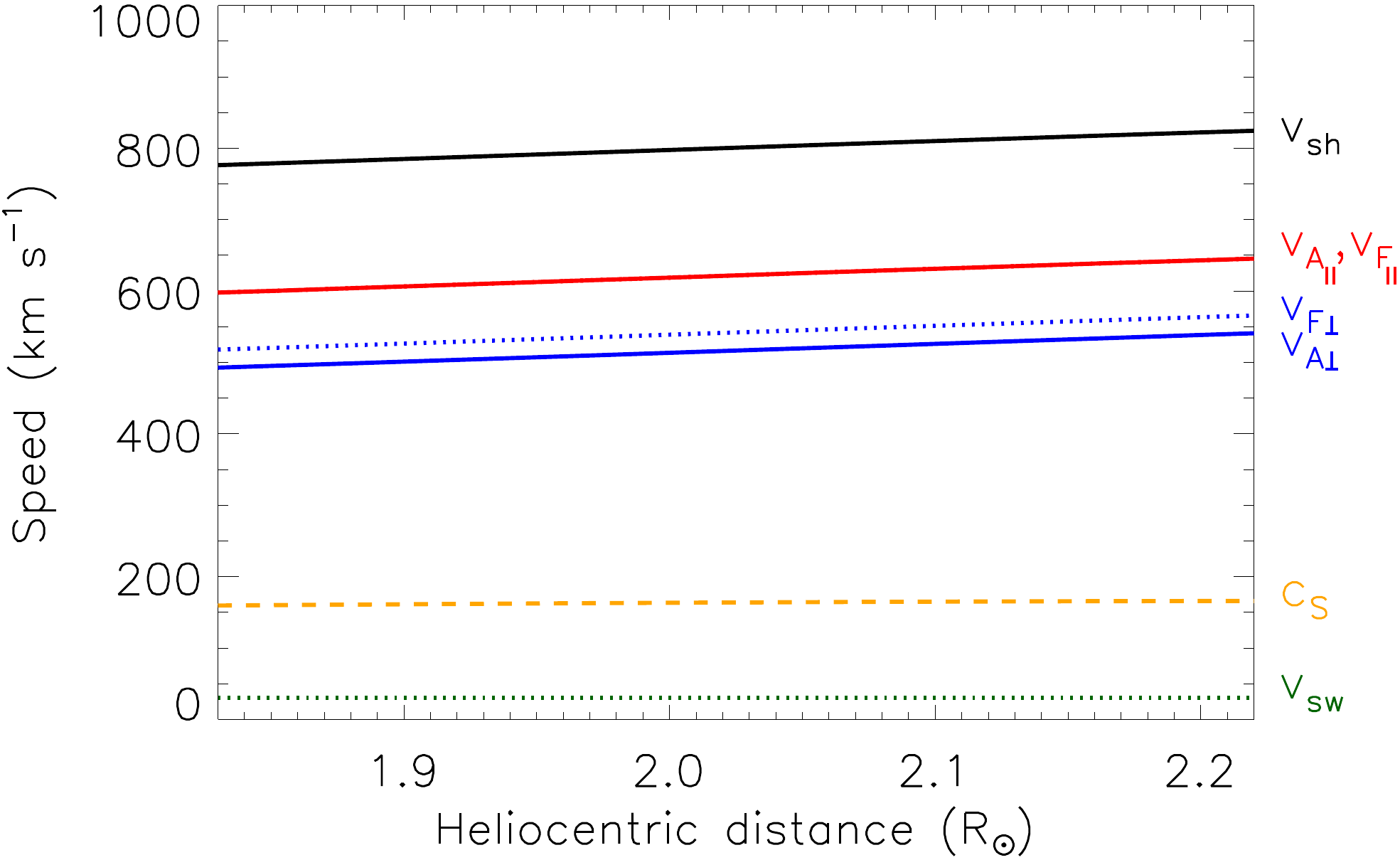}
\caption{Radial profile of the shock speed $v_\mathrm{sh}$ and comparison with the Alfv\'en speed $v_\mathrm{A}$ and the fast-mode speed $v_{\rm F}$ in the ambient corona for both parallel ($\parallel$) and perpendicular ($\perp$) cases. The sound speed  $c_\mathrm{S}$ and the solar wind speed $v_\mathrm{sw}$ are also shown in the same graph. }
\end{figure}

\section{Data analysis and results}

According to MHD theory, the heating and compression of the downstream plasma depend on whether the shock is quasi-perpendicular ($\theta_\mathrm{Bn} > 45\degr$) or quasi-parallel ($\theta_\mathrm{Bn} <  45\degr$), with $\theta_\mathrm{Bn}$ the angle between the upstream magnetic field direction and the shock normal. Assuming that the solar wind is an ideal, adiabatic steady flow of plasma, the fluid equations and Maxwell's equations can be integrated across a MHD shock to give a set of Rankine-Hugoniot jump conditions relating plasma properties on each side of the front. The observed band-split frequencies map the electron densities behind ($n_\mathrm{d}$) and ahead ($n_\mathrm{u}$) of the shock, so that in  the upstream region the plasma emits radio waves at the frequency $f_{\rm L}$, while the compressed plasma in the downstream region ($n_\mathrm{d} > n_\mathrm{u}$) emits radio waves at frequency $f_\mathrm{U} > f_\mathrm{L}$.  In Fig. 3, we show the measured emission frequencies as a function of time as derived from the analysis of the upper ($f_\mathrm{U}$) and lower ($f_\mathrm{L}$) frequency branches of the harmonic band as obtained  from the dynamic spectrum at selected times, together with linear fits to the logarithmic data. The band splitting, simply related to the density jump $X$ at the shock front by $X=n_\mathrm{d}/n_\mathrm{u}=(f_{\rm U}/f_{\rm L})^2$, implies modest (and slowly decreasing) compression ($X \sim 1.5$); $X$  is related to the Alfv\'en Mach number $M_\mathrm{A}$, the plasma-to magnetic pressure ratio  $\beta \equiv 2 c_{\rm S}^2 / \gamma v_{\rm A}^2$, and $\theta_\mathrm{Bn}$  by the cubic equation (e.g., Melrose 1986;  Li \& Cairns 2012)
\begin{equation}
AX^3 + BX^2 + CX + D = 0 ,
\end{equation}
where
\begin{eqnarray}
&& A = - M_\mathrm{A}^2(\gamma - 1) \cos^2{\theta_\mathrm{Bn}} - \beta\gamma\cos^4{\theta_\mathrm{Bn}} ,\nonumber\\
&& B = M_\mathrm{A}^4(\gamma - 2 + \gamma \cos^2{\theta_\mathrm{Bn}}) + M_\mathrm{A}^2(\gamma + 1 + 2\beta\gamma ) \cos^2{\theta_\mathrm{Bn}} ,\nonumber\\
&& C =  - M_\mathrm{A}^6(\gamma - 1) - M_\mathrm{A}^4(\gamma + 2) \cos^2{\theta_\mathrm{Bn}} - M_\mathrm{A}^4\gamma (\beta + 1) ,\nonumber\\
&& D = M_\mathrm{A}^6(\gamma + 1) . \nonumber
\end{eqnarray}
The above are all upstream quantities  ($\gamma = 5/3$ is the ratio of the specific heats). The sound speed profile $c_{\rm S}(r)$ was  calculated as in Gibson et al. (1999), under the assumption that the plasma in the streamer belt is in radial hydrostatic equilibrium, yielding a maximum temperature of $\sim 1.2\cdot 10^6$ K at  2.2 $R_{\sun}$. Foley et al. (2002), who analyzed the radial temperature behavior of coronal streamers at solar minimum by using the CDS/SOHO instrument, obtained a similar result. Given the compression ratio X, the above equation was solved for $M_\mathrm{A}$, with $\beta$ expressed as a function of  $M_\mathrm{A}$. 
Figure 4 graphically shows the results obtained by solving eq. (1) for both quasi-perpendicular and quasi-parallel cases. The strong dependence on $\theta_\mathrm{Bn}$ is clearly evident.  Since $X=X(r)$, we are also able to infer the radial profile of  $M_\mathrm{A}$ for arbitrary  $\theta_\mathrm{Bn}$. Given  $M_\mathrm{A} = M_\mathrm{A}(r,\theta_\mathrm{Bn})$ and an estimate for the shock speed $ v_\mathrm{sh}$, we  obtain the radial dependence of the characteristic plasma speeds $v_\mathrm{A}$ and $v_\mathrm{F}$ (shown in Fig. 5 for the two cases of exactly perpendicular and parallel cases). For $v_\mathrm{sw}$, we adopted the upper limit of the estimate given by Strachan et al. (2000) at solar minimum obtained in the streamer belt  from UVCS/SOHO data ($\lesssim30$ km/s). The difference between  $v_\mathrm{A}$ and $v_\mathrm{F}$ is given by the contribution of the sound speed  $c_{\rm S}$ ($\beta \approx 0.1$ in the perpendicular case). Relying on the above estimate of $v_\mathrm{A}$, we are finally able to infer  $B(r,\theta_\mathrm{Bn})$. The result is shown in Fig. 6. In the same figure, we also show an average estimate obtained by a similar analysis of the band splitting observed  in the WAVES dynamic spectrum around 10 MHz. This analysis, however, is more uncertain owing  to both the very small amount of available data and the higher heliocentric distance ($\sim 3$ $R_{\sun} $) that makes the value of the solar wind speed more uncertain.

\begin{figure}
\centering
\includegraphics[width=0.9\textwidth]{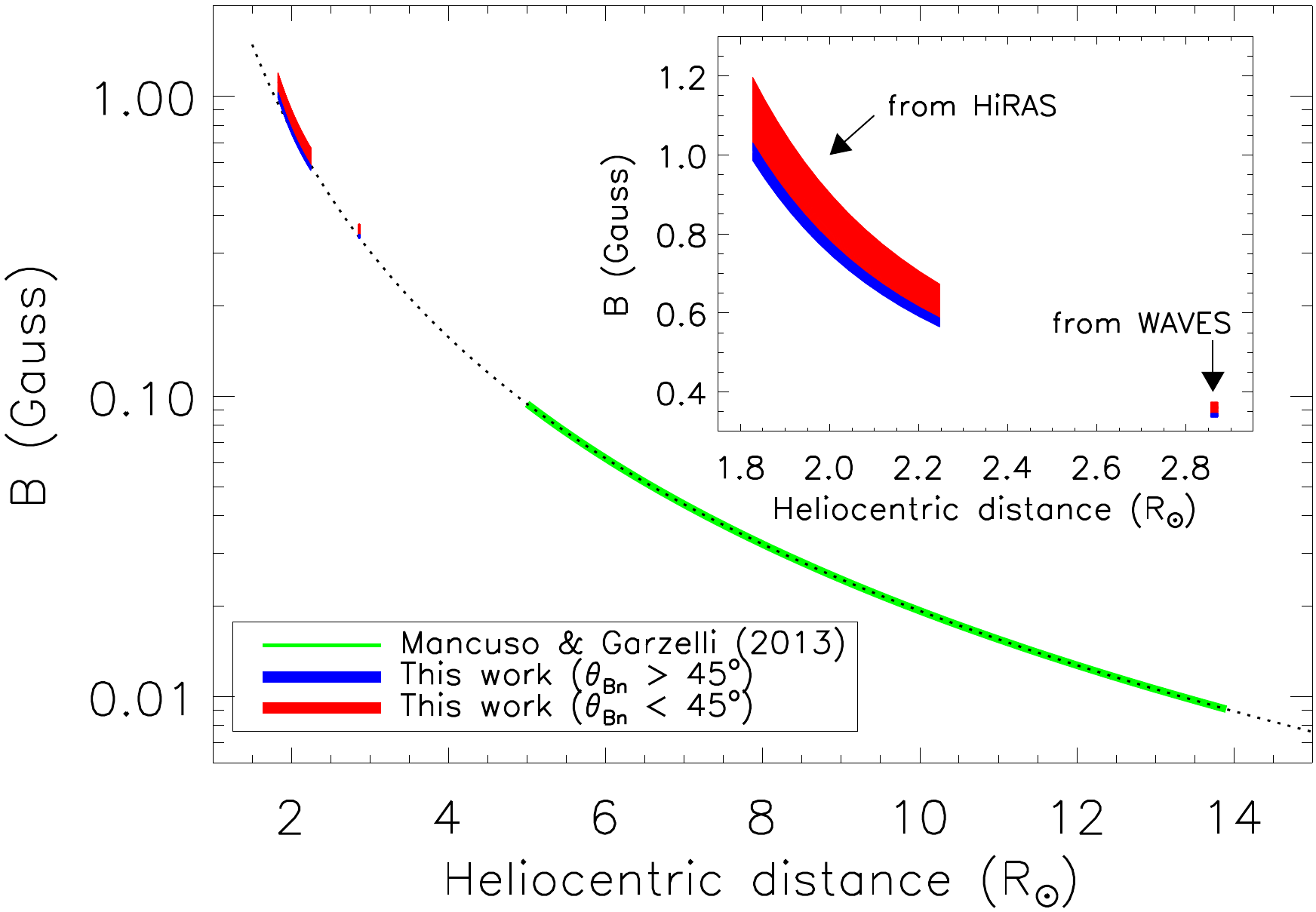}
\caption{Radial profile of the magnetic field strength $B$: results from this work (see inset) and comparison with the profile inferred from Faraday rotation measurements of radio sources occulted by the corona ($ r \approx 5 - 14$ $R_{\sun}$; {\it green line}). Different colors identify the results obtained in the quasi-perpendicular ({\it blue area}) and quasi-parallel ({\it red area}) cases using both HiRAS ($ r \approx 1.8 - 2.2$ $R_{\sun}$) and WIND ($ r \approx 2.8$ $R_{\sun}$) data. The power law of the form $ B(r) ~ = ~ 3.76 ~ r^{-2.29} { ~ ~ \rm G}$ derived by  Mancuso \& Garzelli (2013; {\it dotted line}) nicely matches the two sets of data.}
\end{figure}

\section{Summary and discussion}

We have analyzed the band-splitting of the 1997 May 12 Type~II burst to determine the coronal magnetic field strength  in the heliocentric distance range $ r \approx 1.8 - 2.9$ $R_{\sun}$. The coronal background density $n_\mathrm{e}(r)$  was obtained by inverting white-light $pB$ coronagraph data, and the shock speed was inferred from the Type~II frequency drift. In this work, instead of limiting our study to  perpendicular propagation, we considered arbitrary $\theta_\mathrm{Bn}$, since mechanisms for acceleration of electrons at both quasi-perpendicular and quasi-parallel shocks have been devised (e.g.,  Holman \& Pesses 1983; Mann et al. 2001).  By using the Rankine-Hugoniot relations for MHD, which relate plasma properties on each side of the shock front, and using the compression ratio $X$ inferred by the width of the band split, we found solutions for the Alfv\'en Mach number $M_\mathrm{A}$ for arbitrary values of $\theta_\mathrm{Bn}$. In the quasi perpendicular case $M_\mathrm{A} \approx  1.4 - 1.5$, while in the parallel case $M_\mathrm{A} \approx  1.25$, a value that is probably not sufficient to induce the acceleration of the electron beams responsible for the observed Type~II emission; Type II emission requires $M_\mathrm{A} \gtrsim 1.4$ (Mann et al. 2003). 
 The coronal magnetic field strength $B$ inferred in this work by the analysis of the band-splitting from HiRAS and WAVES radio dynamic spectra ($ r \approx 1.8 - 2.8$ $R_{\sun}$)  for both quasi-perpendicular  and quasi-parallel  cases is shown in Fig. 6. In the same figure, we plot the radial profile obtained higher up in the corona  ($ r \approx 5 - 14$ $R_{\sun}$) by Mancuso \& Garzelli (2013) through  Faraday rotation measurements of extragalactic radio sources occulted by the corona. The best-fit radial profile derived in that work, expressed by the power-law $ B(r) ~ = ~ 3.76 ~ r^{-2.29} { ~ ~ \rm G}$, nicely describes the combined set of data  in a wide range of heliocentric distances ($ r \approx 1.8 - 14$ $R_{\sun}$).

\begin{acknowledgements}
LASCO $pB$ data are produced by a consortium of the NRL, Max-Planck-Institut f\"ur Aeronomie, Laboratoire d'Astronomie Spatiale and Univ. of Birmingham. 
SOHO is a project of international cooperation between ESA and NASA. MLSO is operated by the HAO, a division of the NCAR sponsored by the NSF.
\end{acknowledgements}

\end{document}